# Superionic Fluoride Gate Dielectrics with Low Diffusion Barrier for Advanced Electronics

Kui Meng[1][†], Zeya Li[1][†], Peng Chen[1][†], Xingyue Ma[1][†], Junwei Huang[1], Jiayi Li[1], Feng Qin[1], Caiyu Qiu[1], Yilin Zhang[1], Ding Zhang[2], Yu Deng[1], Yurong Yang[1]*, Genda Gu[3], Harold Y. Hwang[4,5,6], Qi-Kun Xue[2,7]*, Yi Cui[4,5,8]*, Hongtao Yuan[1]*

[1] National Laboratory of Solid State Microstructures, College of Engineering and Applied Sciences, Jiangsu Key Laboratory of Artificial Functional Materials, and Collaborative Innovation Center of Advanced Microstructures, Nanjing University, Nanjing 210000, China.

[2] State Key Laboratory of Low Dimensional Quantum Physics and Department of Physics, Tsinghua University, Beijing 100084, China.

[3] Condensed Matter Physics and Materials Science Department, Brookhaven National Laboratory, Upton, NY 11973, USA.

[4] Stanford Institute for Materials and Energy Sciences, SLAC National Accelerator Laboratory, Menlo Park, CA 94025, USA.

[5] Geballe Laboratory for Advanced Materials, Stanford University, Stanford, CA 94305, USA.

[6] Department of Applied Physics, Stanford University, Stanford, CA 94305, USA.

[7] Department of Physics, Southern University of Science and Technology, Shenzhen 518055, China.

[8] Department of Material Science and Engineering, Stanford University, Stanford, CA 94305, USA.

*Correspondence to: htyuan@nju.edu.cn (H.T.Y.), yicui@stanford.edu (Y.C.), qkxue@mail.tsinghua.edu.cn (Q.K.X.), yangyr@nju.edu.cn (Y.R.Y.).

[†]These authors contributed equally to this work.

Exploration of new dielectrics with large capacitive coupling is an essential topic in modern electronics when conventional dielectrics suffer from the leakage issue near breakdown limit. To address this looming challenge, we demonstrate that rare-earth-metal fluorides with extremely-low ion migration barriers can generally exhibit an excellent capacitive coupling over 20 µF cm$^{-2}$ (with an equivalent oxide thickness of ~0.15 nm and a large effective dielectric constant near 30) and great compatibility with scalable device manufacturing processes. Such static dielectric capability of superionic fluorides is exemplified by MoS$_2$ transistors exhibiting high on/off current ratios over 10$^8$, ultralow subthreshold swing of 65 mV dec$^{-1}$, and ultralow leakage current density of ~10$^{-6}$ A cm$^{-2}$. Therefore, the fluoride-gated logic inverters can achieve significantly higher static voltage gain values, surpassing ~167, compared to conventional dielectric. Furthermore, the application of fluoride gating enables the demonstration of NAND, NOR, AND, and OR logic circuits with low static energy consumption. Notably, the superconductor-to-insulator transition at the clean-limit Bi$_2$Sr$_2$CaCu$_2$O$_{8+\delta}$ can also be realized through fluoride gating. Our findings highlight fluoride dielectrics as a pioneering platform for advanced electronics applications and for tailoring emergent electronic states in condensed matters.

Controlling electronic states in solids with gate dielectrics plays a significant role in condensed matter physics and practical device applications[1-3]. Compared to field-effect transistors based on conventional oxide dielectrics, whose performance is limited by the degree of capacitive coupling, leakage current or dielectric breakdown[4], the electrolyte gating technique has recently been developed as a universal approach to enable very large capacitive coupling at low operation voltage for electronic and energy storage devices[5,6]. Experiments with organic electrolytes have demonstrated that the large capacitive coupling achieved at an electric double-layer (EDL) interface matches the low-voltage operation requirement for device applications and enables high-density carrier accumulation to realize emergent strongly-correlated phenomena in condensed matter[7-9]. However, these organic electrolytes are normally incompatible with present semiconductor manufacturing processes, making electrolyte-based devices difficult to integrate in modern electronics. Therefore, discovering solid-state superionic electrolytes as high-capacitive-coupling dielectrics compatible with manufacturing processes is significant for scalable electronics and integrated circuits but remains challenging.

In this paper, we demonstrated that a catalogue of solid-state superionic fluoride thin films can be used as excellent dielectric materials to address the above-mentioned long-standing issues. We discovered that superionic fluoride dielectric films, taking rare-earth metal fluoride (denoted RE-$F_3$) as an example, exhibit ultrahigh static capacitive performance with capacitance values exceeding 20 μF cm$^{-2}$ (with an equivalent oxide thickness (EOT) of ~0.15 nm) at 10 mHz, comparable to those of organic electrolytes[10]. Notably, the leakage current density of these fluoride dielectrics is as low as 10$^{-6}$ A cm$^{-2}$ when the gate voltage is lower than 3 V. The fluoride-gated $MoS_2$ transistors exhibit high static on/off current ratios of over $10^8$ within the gate voltage ($V_G$) of $\pm 1.0\,\text{V}$ and near-ideal subthreshold swing (SS) of $65\,\text{mV dec}^{-1}$. Significantly, by integrating fluoride-gated $n$-type $MoS_2$ and $p$-type $WSe_2$ transistors, we achieved logic circuits for NOT, NAND, NOR, AND, and OR operations with low static energy consumption. Note that the logic inverters (logic NOT operation) exhibit a high static voltage gain up to 167, exceeding other reported values in inverters based on transition metal dichalcogenides. Using fluorides as gate dielectrics, we found that an electrically-driven superconductor-to-insulator transition in the clean limit[11,12] can be easily obtained in $Bi_2Sr_2CaCu_2O_{8+\delta}$ (Bi-2212), which indicates a large tunability of the fluoride gating on electronic states in quantum materials.

**Dielectric behavior of rare-earth metal fluorides**

A series of rare-earth metal fluoride compounds were chosen as our investigated dielectric materials for the following reasons. First, these RE-$F_3$ are likely superionic conductors because small $F^-$ ions can easily move inside the metal lattice framework[13]. Taking the rare-earth metal fluoride $LaF_3$ (space group $P\bar{3}c1$, Fig. 1a) as an example, the $F^-$ ion sites in the center of the $La^{3+}$ tetrahedrons or octahedrons (defined as $F_T$ or $F_O$ sites) can always serve as the hopping sites for $F^-$ migration. Interestingly, the hopping between $F_T$ sites (defined as the T-T path) and between $F_O$ sites (defined as the O-O path) can provide a low energy barrier for $F^-$ ion migration in the fluoride lattice, resulting in high ionic conductivity particularly under an external electric field (details discussed later). Second, most metal fluorides can form wafer-scale uniform thin films for scalable device fabrication using thermal evaporation. In our case, 4-inch wafer-scale fluoride films (for more than 20 types) with nanometer-scale surface roughness, high crystalline quality, and uniform element distribution can be easily obtained using thermal evaporation (Supplementary Figs. 1 and 2). Third, certain metal fluorides are optical up-conversion materials for two-photon processes[14] or ferromagnetic materials[15], showing great potential to integrate new functionalities into transistors and other electronic devices (Supplementary Figs. 3 and 4, and Supplementary Table 1).

To experimentally demonstrate the excellent capacitive coupling of fluoride dielectrics, we performed electrochemical impedance spectroscopy (EIS) measurements in Au/fluoride/Au sandwiched capacitor structures to obtain the frequency-dependent capacitance. As shown in Fig. 1b, the low-frequency EDL capacitance ($C_{EDL}$) of a representative fluoride film (200-nm-LaF$_3$) reaches 20 µF cm$^{-2}$ at 10 mHz. Such a EDL capacitance value of fluorides is larger than that of the 20-nm HfO$_2$ at the same low-frequency limit (Fig. 1b) and is comparable to those of widely-used liquid-form organic electrolytes[10]. The low-frequency $C_{EDL}$ increases to 100 µF cm$^{-2}$ when a small bias voltage is applied between the top and bottom Au electrode plates (Supplementary Fig. 5), indicating the great capacitive performance in fluoride dielectrics. Such large capacitive coupling makes RE-F$_3$ a competitive dielectric candidate for advanced electronic devices.

Similar to other electrolytes, the obtained capacitance of fluoride dielectrics is frequency-dependent in the EIS measurements within the Au/fluoride/Au sandwiched capacitor structures. Since the EDL charging mechanism has an upper frequency limit ($f_c$) for the time constant ($\tau = 1/f_c = R_{bulk}C_{EDL}$) in the two resistor-capacitor (RC) circuit model[10] for EDL charging speed (where $R_{bulk}$ is the resistance of the electrolyte), two different mechanisms for capacitance coupling in such a sandwiched capacitor can be distinguished in the frequency range from mHz to MHz: Low-frequency EDL capacitance and the high-frequency geometry capacitance between two electrodes. Although the large EDL capacitance coupling is only achieved at low frequency and gradually decreases with increased frequency, which might limit the fast-switching applications of these dielectrics at high frequency, we can actually improve the high-frequency capacitance by thinning down the dielectric thickness (Fig. 1b and Supplementary Fig. 6). For example, the capacitance of 25-nm fluoride-dielectrics at 1 MHz can reach 1 $\mu$F cm$^{-2}$ (corresponding to EOT ~3.5 nm), which is comparable to the capacitance of HfO$_2$ with similar thickness (20 nm) at such a high frequency regime (Fig. 1b).

Excellent dielectric properties in fluoride compounds can be generally modulated with "cation engineering", as discussed below. Figure 1c shows the room-temperature capacitance $C_{EDL}$ of these fluorides, including their crystal structures (Supplementary Fig. 3). Three interesting things need to be addressed here. First, most rare-earth metal fluorides exhibit large capacitive coupling over 10 $\mu$F cm$^{-2}$ (corresponding to the equivalent oxide thickness of ~0.3 nm) at 10 mHz, which can be new members in the dielectric families beyond oxides and nitrides. Interestingly, for RE-F$_3$ fluorides such as LaF$_3$, CeF$_3$, NdF$_3$, SmF$_3$, and EuF$_3$, the $C_{EDL}$ values increase with increasing ionic radius of the corresponding lanthanide metals (Fig. 1d). This provides deep insight into "cation engineering" for improving capacitive coupling: a larger ionic radius enlarges the metal framework and further lowers the barrier height for F$^-$ ion migration inside the lattice, which finally induces a larger ionic conductivity and enhances the $C_{EDL}$ value. Second, the capacitive coupling in fluorides can be further improved by alloy engineering with heterovalent metal cations. For example, the $C_{EDL}$ value of LaF$_3$ film can be doubled by increasing the Sr doping level $x$ (Supplementary Figs. 7 and 8). In mixed La$_{1-x}$Sr$_x$F$_{3-x}$ film that retains the LaF$_3$ hexagonal structure, the defects and domain boundaries (Supplementary Figs. 9 and 10, Supplementary Table 2), introduced by mixing heterovalent metal ions, can always reduce the barrier height of $F^-$ ion migration and further increase the ionic conductivity[16]. Third, cation engineering by introducing magnetic transition metals in fluorides brings emergent magnetic properties to these dielectrics. For example, the transition-metal-based ferromagnetic fluorides $NiF_3$, $YbF_3$, and $GdF_3$ are feasible for magneto-optical devices[15,17] and the ferromagnetic/antiferromagnetic fluorides $FeF_2$ and $MnF_2$ can be used for non-volatile memory devices[18]. Therefore, the fluoride compounds in the "element periodic table" provide a functionalized dielectric family for advanced device applications.

Due to the large EDL capacitive coupling and wide bandgap, rare-earth metal fluoride films exhibit large static dielectric constant ($\varepsilon_{10\ mHz} \sim 30$) and ultra-small EOT with ultralow leakage currents (Supplementary Figs. 11−13). Figure 1e summarizes the EOT at 10 mHz and the leakage current density of our fluorides and other representative dielectric materials[19-26]. Although great efforts have been made in past decades, it is still challenging to achieve low leakage current density and sub-1-nm EOT values simultaneously. For example, high-$\kappa$ $HfO_2$, $Al_2O_3$ and $La_2O_3$ oxide dielectrics[19,23,24] match the requirement for the leakage current of the low-power limit ($< 0.015\ A\ cm^{-2}$), but it is always difficult to achieve sub-1-nm EOT in devices, while layered insulating materials such as thin $h$-BN allow sub-1-nm EOT but always exhibit an extremely high leakage current[21] (higher than $10^{-2}\ A\ cm^{-2}$) at the atomically-thin limit. In sharp contrast, the static EOT values of our fluoride dielectrics (including $CeF_3$, $NdF_3$, $SmF_3$, and $LaF_3$) are much smaller than 1 nm (~0.15 nm at 10 mHz), with the leakage current density maintained at an ultralow level of $10^{-6}\ A\ cm^{-2}$ at gate voltage of 3 V. These results demonstrate that fluorides with large capacitance coupling are excellent dielectric materials for those EDL-based transistors.

To understand low energy barriers of $F^-$ ion migration in fluorides at the atomic scale, which results in the excellent capacitive coupling, we performed first-principles calculations and *ab initio* molecular dynamics (AIMD) simulations. As shown in Fig. 1f–i, the $F^-$ ion migration in the fluoride lattice exhibits extremely-low energy barriers down to ~0.10 eV due to the existence of tetrahedral channel (T1-T2-T3 path) therein (more details in Supplementary Figs. 14–18, Supplementary Table 3). Note that such low energy barriers of $F^-$ ions migration through tetrahedral sites in metal fluorides are even much smaller than those of $Li^+$ cation migration in the well-developed lithium superionic compounds (usually in the range of 0.2~1.0 eV)[27-33] , enabling the highly efficient $F^-$ migration and large capacitance coupling (Fig. 1j). These findings suggest that the extensive tetrahedral cation frameworks in fluoride serve as general low-energy barrier channels for $F^-$ ion migration in these superionic compounds, providing important insight into the understanding of ionic transport in $F^-$ ion conductors and serving as design principles for future discovery of improved solid electrolytes for fluoride dielectrics.

**Fluoride-gated $MoS_2$ transistors**

The superiority of fluoride dielectrics for electronic devices can be demonstrated by fluoride-gated $MoS_2$ transistors with near-ideal SS value, large on/off ratio, small hysteresis, small gate leakage current, low operation voltage, and good reversibility (Fig. 2 and Supplementary Figs. 19–23). On the one hand, even at a small $V_G$ (~1 V), fluoride-gated ($LaF_3$) $MoS_2$ transistors exhibit quite large static on/off current ratios ($I_{on}/I_{off}$) of $3 \times 10^8$ at a source–drain voltage ($V_{DS}$) of 0.7 V (Fig. 2a) with a reasonably-small hysteresis loop ~ 0.1 V ($V_G$ sweep rates between 0.5 V/s to 2 V/s), which is comparable to that of transistors based on layered semiconductors with widely-used high-$\kappa$ dielectric materials[34] such as HfO (more details in Supplementary Fig. 22). On the other hand, the fluoride-gated $MoS_2$ transistors exhibit a static SS value as small as 65 mV dec$^{-1}$ (Fig. 2b), which is comparable to that of $MoS_2$ transistors with high-$\kappa$ dielectric HfO$_2$ (ref. [19]). Such an SS value, approaching the thermionic limit at room temperature (60 mV dec$^{-1}$), permits turning the switching of the transistor within a narrow gate voltage range (threshold voltage ~0.5 V) and illustrates the high quality of the fluoride–$MoS_2$ interfaces. As shown in Fig. 2c, the $I_{on}/I_{off}$ and SS values of our fluoride-gated $MoS_2$ transistors are competitive compared with those of $MoS_2$ transistors employing high-$\kappa$ dielectrics, such as HfO$_2$ and Al$_2$O$_3$ (more details in Supplementary Table 4). More importantly, the maximum switching speed of fluoride-gated transistors can be as fast as 250 ns (Supplementary Figs. 24−27). Such a high on/off ratio and low SS value is a direct result of the excellent EDL capacitive coupling of fluorides, demonstrating the operation capability of fluoride gating for tuning the electronic state in two-dimensional semiconductors and correlated materials.

To evaluate the insulating properties of the fluoride film in practical devices, we track the gate leakage current of the fluoride-gated $MoS_2$ transistors (Fig. 2d). One can see that the fluoride-gated $MoS_2$ transistors show a gate leakage current density (< 10$^{-5}$ A cm$^{-2}$ for gate voltages at 3 V) far below the low-power limit for CMOS devices[35]. The extremely low leakage current with an ultralow static EOT value as small as 0.15 nm (at 10 mHz) indicates that rare-earth-metal fluoride dielectrics are sufficiently insulating to minimize the leakage current for electronic devices and integrated circuits. Interestingly, these fluoride-gated transistors show rather good reversibility, reproducibility and thermal stability of device performance (Fig. 2e, Supplementary Figs. 28−32). For example, the on-state and off-state current remain almost the same and the device exhibits an unchanged on/off ratio of ~$10^8$ even after over 500 cycles of switching operations at a frequency of 100 Hz. Our fluoride-gated transistors, which exhibit near-ideal SS value, large on/off ratio, small gate leakage current, low operation voltage, and good reversibility, show promising potential for low-power-advanced electronic applications.

**Fluoride-gated van der Waals logic circuits**

Based on the fluoride-gated MoS$_2$ transistors mentioned above, we further integrated $n$-type MoS$_2$ and $p$-type WSe$_2$ transistors to construct CMOS inverter devices (Fig. 3a). The output voltage signal ($V_{\text{OUT}}$) is step-responsive to the input voltage signal ($V_{\text{IN}}$) under different preset voltages ($V_{\text{DD}}$) from 2.0 to 2.8 V (Fig. 3b). The DC voltage gain is an important figure of merit for CMOS inverters, which is calculated from the slope (d$V_{\text{OUT}}$/d$V_{\text{IN}}$). The DC voltage gain reaches as high as 167 at $V_{\text{DD}}$ = 2.6 V (Fig. 3c), which is the highest value among similar inverters based on transition metal dichalcogenides reported[36–42] so far (Fig. 3d). Correspondingly, the maximum power consumption at even the largest $V_{\text{DD}}$ = 3.0 V (Supplementary Fig. 33) is lower than 140 nW, which gives a low static power level for circuit applications, distinct from those of semiconducting transition-metal-dichalcogenide-based devices with oxide dielectrics (Supplementary Table 5).

To confirm the great noise tolerance in our fluoride-gated inverters, we plot the bistable hysteresis voltage transfer characteristics at $V_{\text{DD}}$ = 2.0 V (Fig. 3e) and extract the noise margins of $NM_{\text{L}}$ = 1.0 V ($NM_{\text{L}}$ is defined as the value of the maximum input low-level voltage minus the maximum output low-level voltage, Supplementary Fig. 33), and $NM_H = 0.8$ V ($NM_H$ is defined as the value of the minimum output high-level voltage minus the minimum input high-level voltage). The total noise margin, $(NM_L + NM_H)/V_{DD}$, always exceeds 90% for all applied $V_{DD}$ (Fig. 3f), indicating a high noise tolerance of the fluoride-gated inverters. As shown in Fig. 3g, the $V_{OUT}-V_{IN}$ response is measured by applying a pulsed square wave $V_{IN}$ with an amplitude of 2.0 V. Interestingly, the inverter operates with a good performance at 100 Hz, 1 kHz and 10 kHz (details in Supplementary Fig. 27). Specifically, for input frequency of 10 kHz, the inverter can still have a response time as fast as 13 μs. Compared to the reported inverters based on transition metal dichalcogenides (Supplementary Table 6), our fluoride-gated inverters exhibit the highest DC voltage gain, excellent noise tolerance, and low static energy consumption.

More importantly, these fluoride-gated transistors can be further integrated into logic circuits. As shown in Fig. 4a, several fluoride-gated transistors based on $p$-type $WSe_2$ (denoted as TP) and $n$-type $MoS_2$ (denoted as TN) are connected in parallel or in series between the power supply and the ground, integrating into logic NAND, NOR, AND and OR gates. Correspondingly, the results of logical operations of these logic gates are given in the truth table in Fig. 4b. Specifically, we denote the applied input low voltage 0 V as logic "0" and high voltage 1.5 V as logic "1", and correspondingly we denote the measured output low and high voltage values as logic "0" and "1". For each logic gate, there are four logic combinations of the input signal, namely, "0-0", "0-1", "1-0" and "1-1". Taking the NAND logic circuit as an example, the two $p$-$WSe_2$ transistors (TP1 and TP2) are connected in parallel to the power supply, and the two $n$-$MoS_2$ transistors (TN1 and TN2) are connected in series to the ground.

As shown in Fig. 4c, the input signals $V_{IN1}$ and $V_{IN2}$ are applied onto the gates of those transistors to control their on/off states, while the output signal $V_{OUT}$ measures the voltage drops on the series-connected TN1 and TN2.

One can see that if either one of the two input signals $V_{IN1}$ and $V_{IN2}$ is the logic "0", the $p$-WSe$_2$ transistors TP1 and TP2 will operate as on states, while $n$-MoS$_2$ transistors TN1 and TN2 will operate as off states. Thus, the $V_{OUT}$ will be at the high-level voltage and this logic gate outputs the logic "1". In sharp contrast, only in the case of both input signals $V_{IN1}$ and $V_{IN2}$ being the logic "1" will the $p$-WSe$_2$ transistors TP1 and TP2 operate as off states, while the $n$-MoS$_2$ transistors TN1 and TN2 operate as on states. Thus, the $V_{OUT}$ will be at the low-level voltage, and this logic gate outputs the logic "0". Such logic operations are shown by the sequence of applying different input signals (Fig. 4d), where each cycle contains four logic combinations. Similarly, a NOR logic gate can be realized by combining two series-connected $n$-MoS$_2$ transistors and parallel-connected $p$-WSe$_2$ transistors (Fig. 4e). Interestingly, based on the logic gates of NAND and NOR, the AND and OR logic gates can be realized through serial connection with a NOT logic gate (Fig. 4g-j). One can see that the logic AND gate outputs logic "1" only in the input logic combination of "1-1", and outputs logic "0" for the rest of the input logic combinations of "0-0", "0-1" and "1-0". Correspondingly, the logic OR gate outputs logic "0" only in the input logic combination of "0-0" and outputs logic "1" for the rest of the input logic combinations of "0-1", "1-0" and "1-1". Note that the power consumptions of these logic circuits are all below 0.1 µW, demonstrating the capability of the fluoride dielectric for logic device integration with low static energy consumption.

**Fluoride-gating induced superconductor-insulator transition**

To further examine the capability of fluoride dielectrics for tuning the electronic states in quantum materials, we fabricated fluoride-gated Bi-2212 devices and tuned the superconductivity therein. Figure 5a,b shows the schematic device geometry and the cross-sectional scanning transmission electron microscopy image of the sharp LaF$_3$/Bi-2212 interface. Once a positive gate voltage is applied, the electrical potential drives the F$^-$ ions away from LaF$_3$/Bi-2212 interface and leads to the "enrichment" of positively charged La$^{3+}$ ions at the interface, which generates the local EDL electric field and induces electron accumulation (hole depletion) in Bi-2212. As a direct result, a superconductor-to-insulator transition with reduced superconducting critical temperature ($T_c$) is achieved without any chemical reaction or chemical intercalation since chemically-active F$^-$ ions has already been driven away from Bi-2212 surface, as shown in Fig. 5c. This transition can be presented more obviously when plotting the normalized resistance $R_{xx}/R_{xx}(200 \text{ K})$ as a function of $T$ and $p$ (Fig. 5d). One can see that the $T_c$ continuously decreases as the $p$ level decreases until the superconductivity is suppressed, which is consistent with the phenomenon in the monolayer Bi-2212 case with in-situ annealing method[22]. Considering that the superconductor-insulator transition requires the hole depletion over the whole flake (~30 nm), we believe that, the giant local EDL electric field can lead to the loss of oxygen in the Bi-2212 flake and further makes the flake insulating, as also reported in CeO$_2$ gated Bi-2212 samples[43]. These electrically-driven superconductor-to-insulator transitions in Bi-2212 provide great opportunities to modulate quantum critical phenomena in high-temperature superconductors[44–46].

Figure 5e shows the temperature-dependent $R_{xx}$ in the critical region of the superconductor-to-insulator transition. All the $R_{xx}-p$ curves at different temperatures crossed at the same doping level $p = 0.021$, corresponding to a clear separatrix for the critical doping level $p_c$. If we scale the horizontal axis as $|p - p_c|T^{-1/zv}$ (where $zv$ is the critical exponent), all the $R_{xx}-p$ curves at different temperatures collapse to one universal curve, where the exponent $zv$ is fitted to be 1.47 (Fig. 5f), which is very close to the clean limit of the superconductor-insulator transition, verifying the clean charge-doping scenario in fluoride-gated Bi-2212 flakes. Such an observation demonstrates that fluoride gating can serve as a clean tool for tuning the intrinsic electronic state and exploring emergent correlated physics without being bothered by percolation and inhomogeneity effects.

**Conclusions**

In summary, we demonstrate a fluoride-based superionic dielectric family with high capacitive coupling for device applications and exemplify its excellent gating capability for tuning the electronic states of semiconductors and quantum materials. We elucidate the general $F^-$ ion migration mechanism with extremely-low energy barriers at the atomic level, which can be used as general guidance for searching large-capacitive-coupling dielectric materials. The current maximum switching speed of fluoride-gated transistors is around 4 MHz, which can be greatly enhanced after device optimization such as reducing the parasitic capacitance between gate and source-drain electrodes, decreasing channel length and source-drain contact resistance, as well as increasing carrier mobilities of channel materials. The wafer-scale fluoride dielectric films and their capability to integrate with the conventional semiconductor lithography process provide a new material platform to develop functional devices and manufacture advanced electronics especially for neuromorphic transistors and energy storage devices.

**Acknowledgements**

This work was supported by grants from the National Natural Science Foundation of China (grant numbers 92365203 (H.T.Y.) and 52072168 (H.T.Y.)) and the National Key R&D Program of China (grant number 2021YFA1202901 (J.W.H)). The work at BNL was supported by grants from the US Department of Energy, office of Basic Energy Sciences (grant number DOE-sc0012704 (G.D.G.)). Y.C. and H.Y.H. acknowledge the support by the U.S. Department of Energy, Office of Basic Energy Sciences, Division of Materials Sciences and Engineering (grant number DE-AC02-76SF00515). The authors also would like to thank Yicheng Shen and Zhenliang Liu for their assistance on electrical transport measurements.

**Author contributions**

K.M., Z.Y.L., P.C. and X.Y.M. contributed equally to this work. H.T.Y., Q.K.X., Y.C. and H.Y.H. conceived and designed the experiments. K.M., Z.Y.L., P.C. and Y.L.Z. performed device fabrications. K.M. and P.C. performed the EIS measurements. K.M., F.Q., D.Z. and J.W.H. performed electrical transport measurements. C.Y.Q. performed AFM measurements. G.D.G. provided high-quality crystals. J.Y.L. and Y.D. performed STEM characterization. X.Y.M. and Y.R.Y. performed theoretical calculations. K.M., Z.Y.L. and F.Q. analysed the data. K.M., Z.Y.L. and H.T.Y. wrote the manuscript with input from all authors.

**Competing interests:** The authors declare that they have no competing interests.

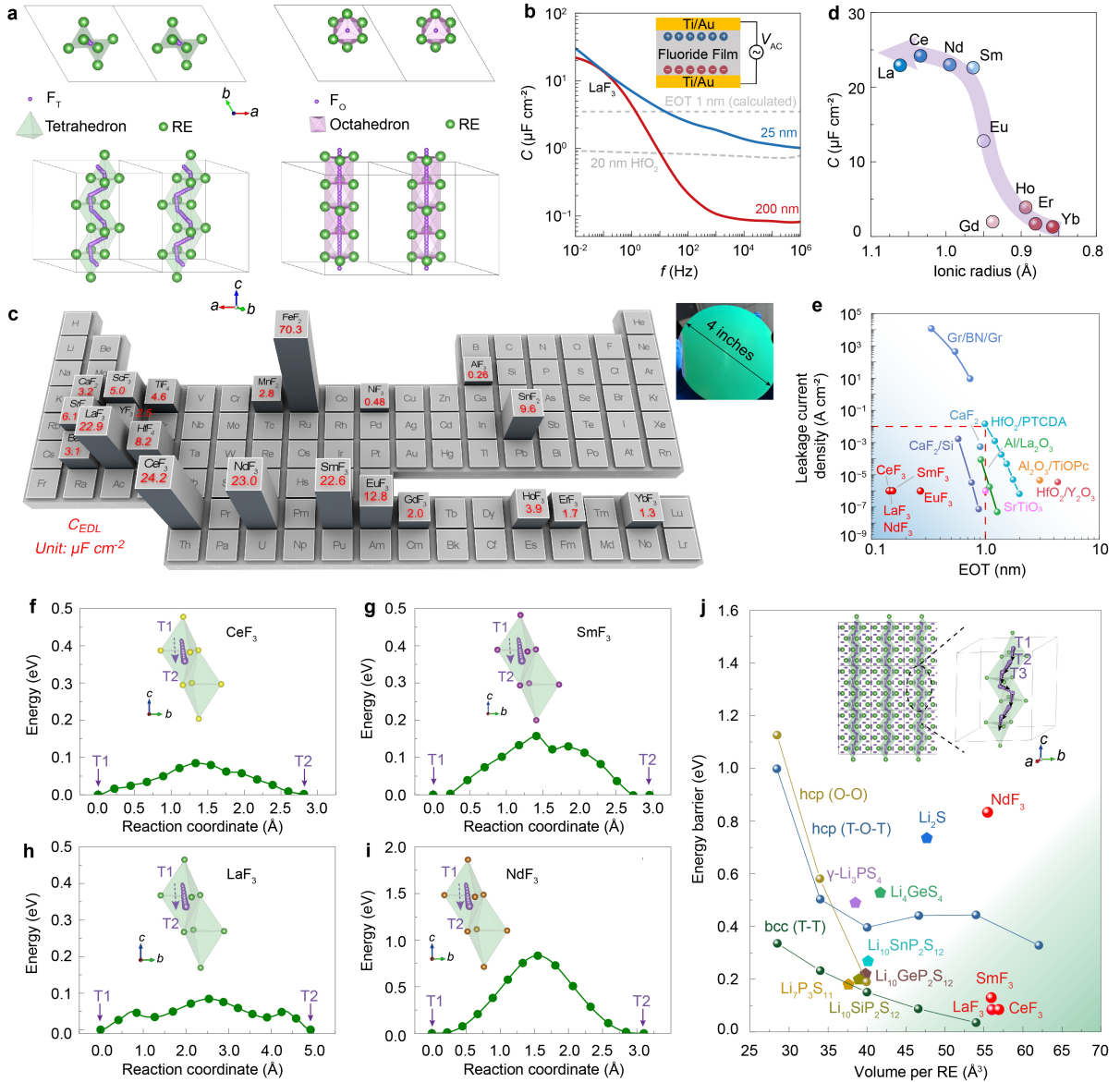

**Fig. 1 | Crystal structure, dielectric properties and F⁻ ion migration for a fluoride dielectric catalogue.**

**a**, The tysonite structure of rare-earth metal fluoride. There are two possible migration paths for F⁻ ion diffusion through adjacent tetrahedral sites (left panel, T-T path) or octahedral sites (right panel, O-O path). **b**, Frequency-dependent capacitance of the parallel plate capacitor based on LaF₃ film (25 nm and 200 nm). Dashed lines are frequency-dependent capacitance of 20-nm HfO₂ and the theoretical capacitance for EOT of 1 nm. Inset: Schematic sandwiched structures of Au/LaF₃/Au. **c**, Metal fluorides shown as an "element periodic table". The $C_{EDL}$ values at 10 mHz are represented by the height of the column n units of $\mu F\ cm^{-2}$. Inset: Optical images of wafer-sized high-quality $LaF_3$ film on 4-inch $Si/SiO_2$ substrates. **d**, Capacitance values for rare-earth lanthanide fluorides as a function of the cation radius. **e**, Experimental leakage current density versus EOT measured with the Au/fluoride/Au structure. Red balls represent data for our fluoride dielectrics ($LaF_3$, $CeF_3$, $NdF_3$, $SmF_3$ and $EuF_3$), while the others represent data for other dielectrics adapted from the literature (refs. [19-26]). The EOT values of our fluoride dielectrics are estimated by their capacitance at 10 mHz. The red dotted rectangle highlights the region for EOT less than 1 nm and leakage current density less than $10^{-2}\ A\ cm^{-2}$.

**f-i**, Illustration of the $F^-$ ion migration path between two face-sharing tetrahedral interstices

(T1-T2, inset) and the corresponding energy barrier for $CeF_3$ (**f**), $SmF_3$ (**g**), $LaF_3$ (**h**) and $NdF_3$

(**i**). Purple balls represent $F_T$ atoms, and other colored balls represent rare-earth metal atoms. **j**,

The energy barrier calculated for the $F^-$ ion migration pathways (T1-T2) in rare-earth metal fluoride lattices and lithium superionic compounds at different lattice volumes (refs. [27-33]). The balls represent the theoretical calculation results corresponding to different structures (hcp and bcc) and paths (O-O, T-O-T, and T-T) in lithium superionic conductors, and the pentagon represents the experimental results in different lithium superionic conductors. Inset: Illustration of the $F^-$ ion migration path in the rare-earth metal fluoride lattice structures.

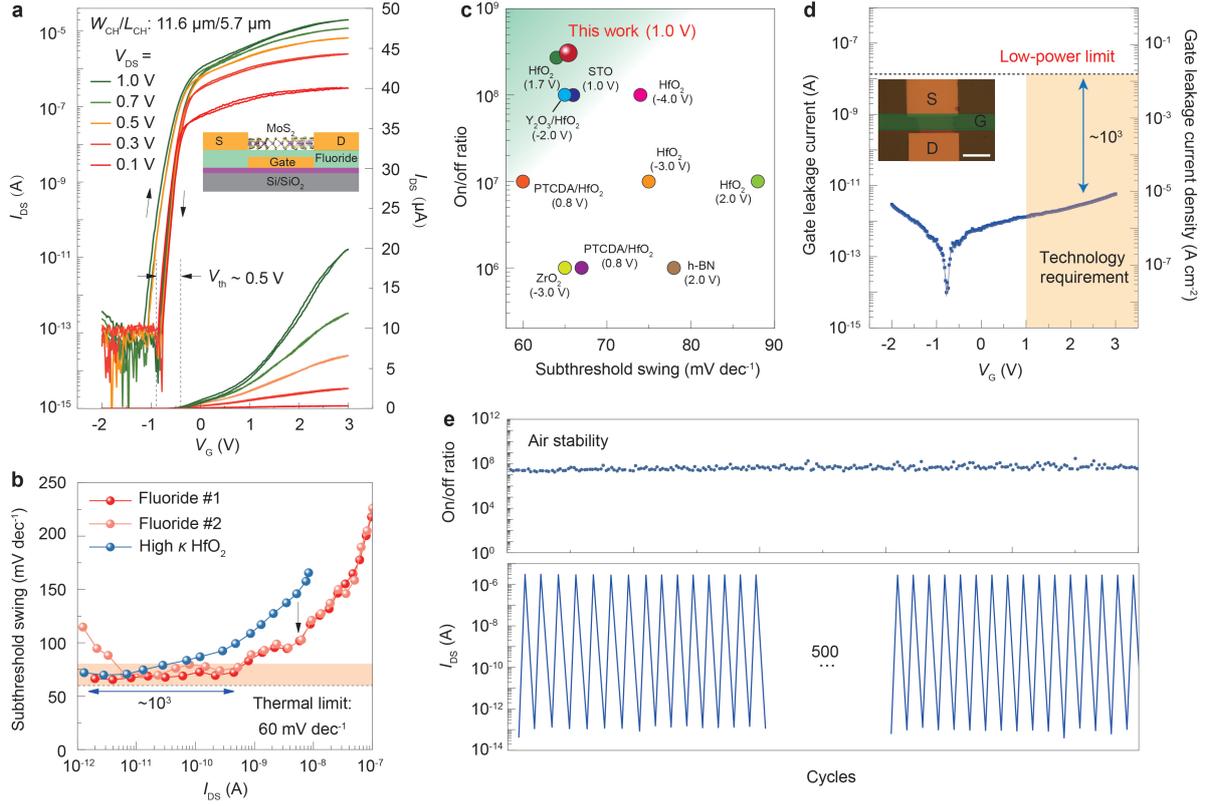

**Fig. 2 | Fluoride-gated MoS₂ transistors.**

**a**, Room-temperature static transfer characteristics (sweeping rate of 100 mV/s) of the $n$-type MoS₂ transistors at different $V_{DS}$. Inset: Schematic of the fluoride-gated MoS₂ transistor. The fluoride (LaF₃) thickness is ~100 nm. **b**, Subthreshold swing (SS) as a function of $I_{DS}$. The dashed line represents the thermal limit of 60 mV dec⁻¹. **c**, A comparison of the current on/off ratio and SS of our fluoride-gated transistor and other MoS₂ transistors with high $\kappa$ dielectrics in the literature (more details in Supplementary Table 4). **d**, Gate leakage current density of MoS₂ transistors. The dashed line represents the low-power limit of 0.015 A cm⁻². The gate leakage current density of our fluoride-gated transistor could be negligible relative to low-power limit. Inset: Optical microscopy image of the fluoride-gated transistor. The scale bar is 10 μm. **e**, On/off ratio (top panel) and switching operation (bottom panel) at a frequency of 100 Hz for fluoride-gated MoS₂ transistors with 500 on/off cycles.

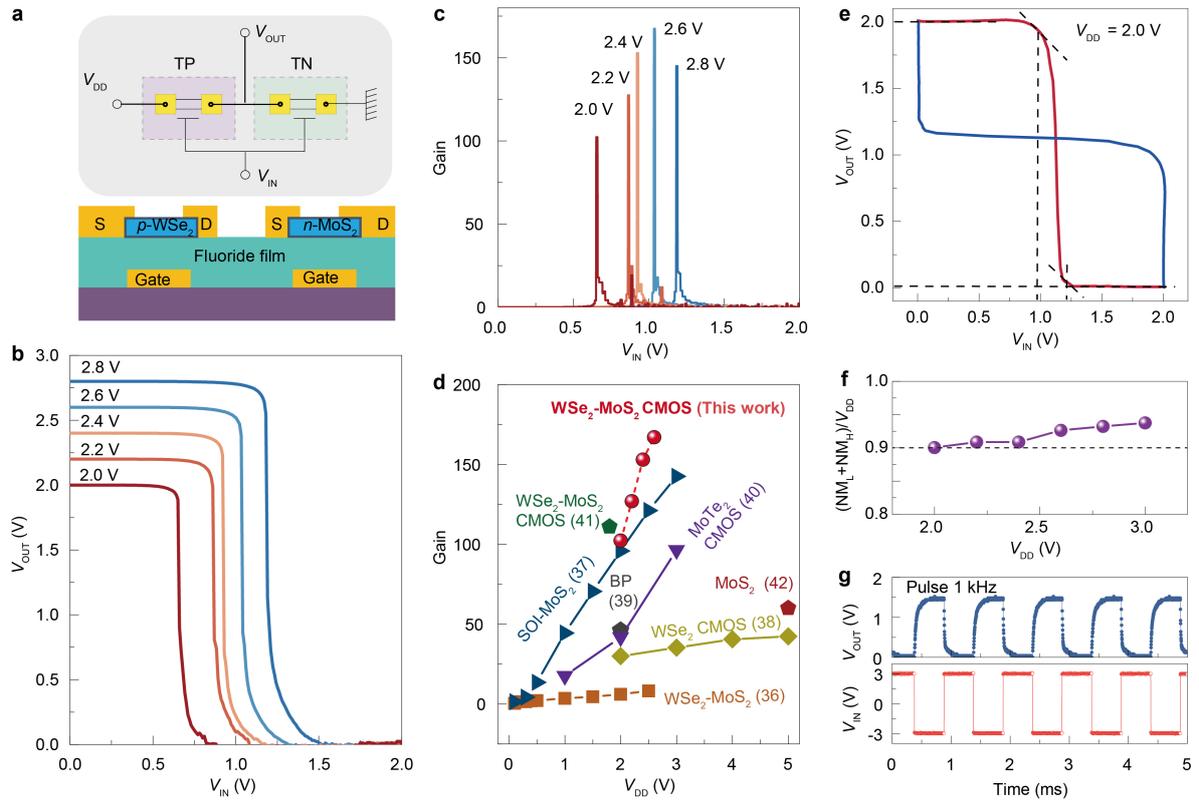

**Fig. 3 | CMOS inverter based on *n*-type MoS₂ and *p*-type WSe₂ transistors.**

**a**, Top panel: equivalent circuit of the inverter. TP and TN represent the *p*-type transistor and *n*-type transistor, respectively. Bottom panel: Illustration of a CMOS inverter based on *n*-type MoS₂ and *p*-type WSe₂ transistors with a fluoride gate. The static gate voltage is used as the input signal, and the common terminal voltage of the two transistors is used as the output signal. The fluoride (LaF₃) thickness is ~100 nm. **b**, Voltage transfer characteristics of the MoS₂/WSe₂ inverter under different $V_{DD}$ values from 2.0 to 2.8 V with a step of 0.2 V (sweeping rate of 100 mV/s). **c**, $dV_{OUT}/dV_{IN}$ as a function of $V_{IN}$. The static voltage gain calculated from the maximum of $dV_{OUT}/dV_{IN}$ at different $V_{DD}$. **d**, Comparison of the static voltage gains between fluoride-gated inverters and other inverters based on transition metal dichalcogenides (refs. [36–42]). **e**, Bistable hysteresis voltage transfer characteristics of the fluoride-gated inverter at $V_{DD}$ =

2.0 V. Two black tangent lines on the $V_{OUT}$–$V_{IN}$ curve represent the value of −1 for $dV_{OUT}/dV_{IN}$, from which value we can determine the $V_{IL}$ and $V_{IH}$ voltages (the abscissa corresponding to the vertical dotted line). The vertical coordinates corresponding to the horizontal dotted line are the $V_{OL}$ and $V_{OH}$ voltages. **f**, The ratio of the total noise margin, $(NM_L + NM_H)/V_{DD}$, as a function of $V_{DD}$. **g**, $V_{OUT}$–$V_{IN}$ response by applying a square-wave $V_{IN}$ with a frequency of 1 kHz and an amplitude of 2.0 V. The fluoride (LaF$_3$) thickness for this inverter is ~50 nm.

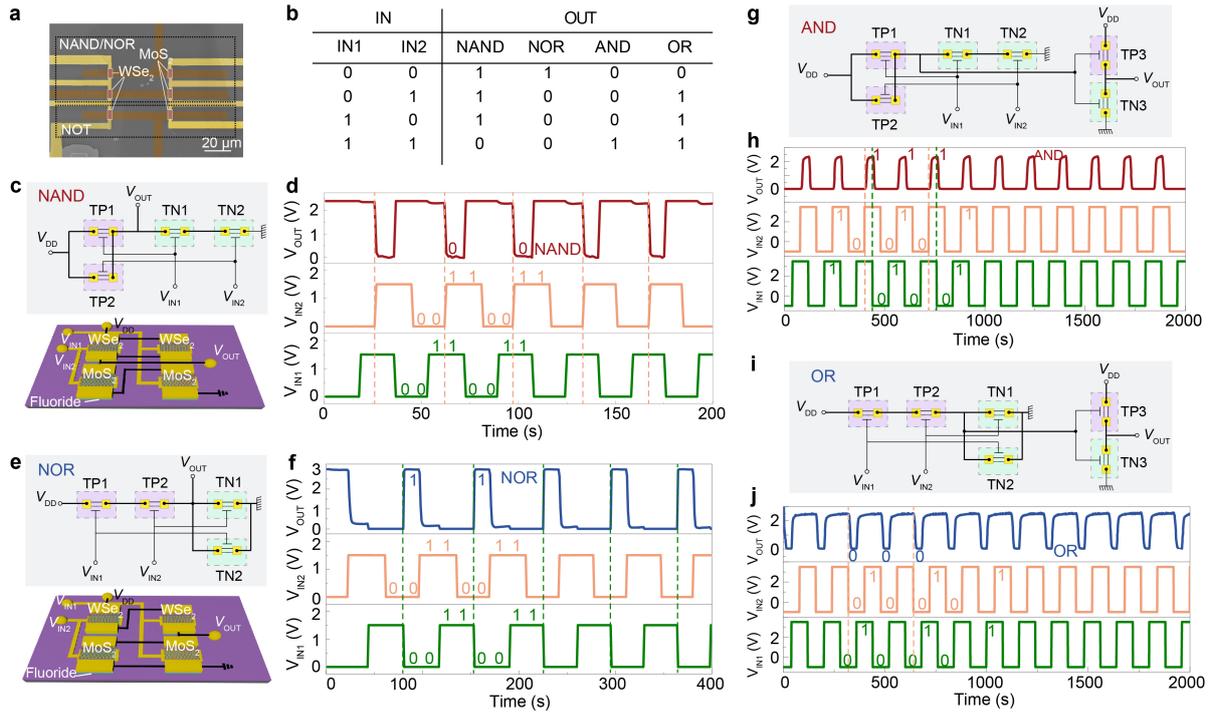

**Fig. 4 | Linear logic gates based on fluoride-gated *n*-type MoS₂ and *p*-type WSe₂ transistors.**

**a**, Scanning electron microscope image of the linear logic gate containing six fluoride-gated transistors. The four transistors in the top black dotted rectangle can form logic NAND and NOR gates, which, together with the inverter (constructed by the other two transistors, bottom black dotted rectangle), can further form logic AND and OR gates. The fluoride (LaF₃) thickness is ~100 nm. **b**, Truth table of logic NAND, NOR, AND, and OR gates. **c-f**, Circuit connection diagram and schematic of the logic gate structure of logical NAND (**c**) and NOR (**e**). The NAND and NOR logic circuit is constructed with two *p*-type WSe₂ (denoted as TP1 and TP2) and two *n*-type MoS₂ transistors (denoted as TN1 and TN2). The output results of logic NAND (**d**) and NOR gates (**f**) by applying a quasi-static square-wave $V_{IN}$ with an amplitude of 1.5 V. **g-j**, Circuit connection diagram of the logic gate structure of logical AND (**g**) and OR (**i**). Based on the logic function of NAND (and NOR), the logic function of AND (and OR) can be realized through the parallel connection with a logic NOT gate. The output results of logic AND (**h**) and OR gates (**j**) by applying a quasi-static square-wave $V_{IN}$ with an amplitude of 3.5 V.

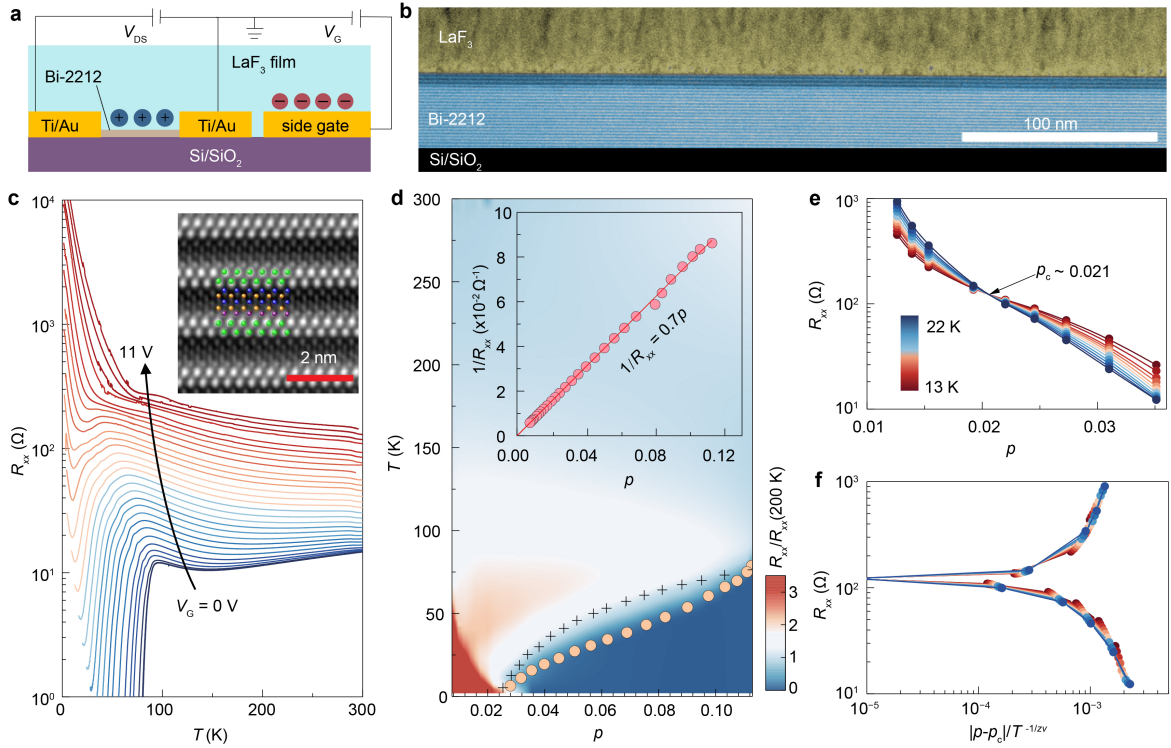

**Fig. 5 | Continuous tuning of the superconductor-to-insulator transition in Bi-2212 with fluoride gating.**

**a**, Schematic illustration of a fluoride-gated device with a side-gate electrode. **b**, False-colored STEM image of the sharp interface between LaF$_3$ film (shaded in yellow) and Bi-2212 (shaded in blue). The scale bar is 100 nm. **c**, Temperature-dependent $R_{xx}$ of the fluoride-gated device. Bi-2212 (~30 nm) is initially underdoped with an $T_c$ of 78 K. $T_c$ is defined at the temperature with the maximum d$R_{xx}$/d$T$. The fluoride (LaF$_3$) thickness is ~200 nm. Inset: schematic atomic lattice overlapped on top of the atomic resolution STEM image of Bi-2212. The scale bar is 2 nm. **d**, Colored mapping of resistance as a function of temperature and doping level. The resistance at each doping level is normalized by its value at 200 K. The orange circles denote the $T_c$ values at each doping level. The hole doping level $p$ is determined by the empirical relation with $T_c = T_{c,max}[1 - 82.6(p - 0.16)^2]$ in the superconducting regime[46]. For comparison, the starting point of the superconducting dome is consistent with reported values in the monolayer case (indicated by the black crosses adapted from ref. [11]), showing a similar doping level for the superconductor-to-insulator transition. Inset: $1/R_{xx}$ as a function of the estimated doping level $p$ with a linear relation of $1/R_{xx} = 0.7p$, in which the $R_{xx}$ values are taken at 200 K in the superconducting regime (indicated by the black dashed interval), and the linear relation can be further extrapolated to the insulating regime to estimate the doping level $p$. **e**, $R_{xx}$ as a function of $p$ at different temperatures. The critical doping level $p_{\mathrm{c}}$ is 0.021. **f**, Scaling analysis of $R_{xx}$ as a function of $|p - p_{\mathrm{c}}|T^{-1/zv}$ near the superconductor-insulator transition, where the $zv$ value is estimated to be 1.47 within a clean limit regime.

**References:**

1.  Kingon, A. I., Maria, J.-P. & Streiffer, S. K. Alternative dielectrics to silicon dioxide for memory and logic devices. *Nature* **406**, 1032–1038 (2000).

2.  Caviglia, A. D. *et al.* Electric field control of the $LaAlO_3/SrTiO_3$ interface ground state. *Nature* **456**, 624–627 (2008).

3.  Cheema, S. S. *et al.* Ultrathin ferroic $HfO_2$–$ZrO_2$ superlattice gate stack for advanced transistors. *Nature* **604**, 65–71 (2022).

4.  Alam, M. A., Smith, R. K., Weir, B. E. & Silverman, P. J. Uncorrelated breakdown of integrated circuits. *Nature* **420**, 378–378 (2002).

5.  Cho, J. H. *et al.* Printable ion-gel gate dielectrics for low-voltage polymer thin-film transistors on plastic. *Nat. Mater.* **7**, 900–906 (2008).

6.  Wang, X. *et al.* Electrode material–ionic liquid coupling for electrochemical energy storage. *Nat. Rev. Mater.* **5**, 787–808 (2020).

7.  Saito, Y., Kasahara, Y., Ye, J., Iwasa, Y. & Nojima, T. Metallic ground state in an ion-gated two-dimensional superconductor. *Science* **350**, 409–413 (2015).

8.  Li, L. J. *et al.* Controlling many-body states by the electric-field effect in a two-dimensional material. *Nature* **534**, 185–189 (2016).

9.  Leighton, C. Electrolyte-based ionic control of functional oxides. *Nat. Mater.* **18**, 13–18 (2019).

10. Yuan, H. *et al.* Electrostatic and electrochemical nature of liquid-gated electric-double-layer transistors based on oxide semiconductors. *J. Am. Chem. Soc.* **132**, 18402–18407 (2010).

11. Yu, Y. et al. High-temperature superconductivity in monolayer $Bi_2Sr_2CaCu_2O_{8+\delta}$. *Nature* **575**, 156–163 (2019).

12. Bollinger, A. T. *et al.* Superconductor–insulator transition in $La_{2-x}Sr_xCuO_4$ at the pair quantum resistance. *Nature* **472**, 458–460 (2011).

13. Wu, C.-L. *et al.* Gate-induced metal–Insulator transition in $MoS_2$ by solid superionic conductor $LaF_3$. *Nano Lett.* **18**, 2387–2392 (2018).

14. Zhou, B., Shi, B., Jin, D. & Liu, X. Controlling upconversion nanocrystals for emerging applications. *Nat. Nanotechnol.* **10**, 924–936 (2015).

15. Chen, Y.-C. *et al.* A brilliant cryogenic magnetic coolant: magnetic and magnetocaloric study of ferromagnetically coupled $GdF_3$. *J. Mater. Chem. C* **3**, 12206–12211 (2015).

16. Motohashi, K., Nakamura, T., Kimura, Y., Uchimoto, Y. & Amezawa, K. Influence of microstructures on conductivity in tysonite-type fluoride ion conductors. *Solid State Ion.* **338**, 113–120 (2019).

17. Mattsson, S. & Paulus, B. Density functional theory calculations of structural, electronic, and magnetic properties of the $3d$ metal trifluorides $MF_3$ (M = Ti-Ni) in the solid state. *J. Comput. Chem.* **40**, 1190–1197 (2019).

18. Higuchi, T. & Kuwata-Gonokami, M. Control of antiferromagnetic domain distribution via polarization-dependent optical annealing. *Nat. Commun.* **7**, 10720 (2016).

19. Li, W. *et al.* Uniform and ultrathin high-$\kappa$ gate dielectrics for two-dimensional electronic devices. *Nat. Electron.* **2**, 563–571 (2019).

20. Illarionov, Y. Y. et al. Ultrathin calcium fluoride insulators for two-dimensional field-effect transistors. *Nat. Electron.* **2**, 230–235 (2019).

21. Britnell, L. *et al.* Electron tunneling through ultrathin boron nitride crystalline barriers. *Nano Lett.* **12**, 1707–1710 (2012).

22. Vexler, M. I., Illarionov, Y. Y., Suturin, S. M., Fedorov, V. V. & Sokolov, N. S. Tunneling of electrons with conservation of the transverse wave vector in the Au/CaF$_2$/Si(111) system. *Phys. Solid State* **52**, 2357–2363 (2010).

23. Iwai, H. *et al.* Advanced gate dielectric materials for sub-100 nm CMOS. *Digest. International Electron Devices Meeting*, 625–628 (2002).

24. Wang, X. *et al.* Improved integration of ultra-thin high-$\kappa$ dielectrics in few-layer MoS$_2$ FET by remote forming gas plasma pretreatment. *Appl. Phys. Lett.* **110**, 053110 (2017).

25. Huang, J.-K. *et al.* High-$\kappa$ perovskite membranes as insulators for two-dimensional transistors. *Nature* **605**, 262–267 (2022).

26. Zou, X. *et al.* Interface engineering for high-performance top-gated MoS$_2$ field-effect transistors. *Adv. Mater.* **26**, 6255–6261 (2014).

27. Wang, Y., Richards, W., Ong, S. *et al.* Design principles for solid-state lithium superionic conductors. *Nat. Mater.* **14**, 1026–1031 (2015).

28. Kuhn, A., Duppel, V. & Lotsch, B. V. Tetragonal Li$_{10}$GeP$_2$S$_{12}$ and Li$_7$GePS$_8$—exploring the Li ion dynamics in LGPS Li electrolytes. *Energy Environ. Sci.* **6**, 3548–3552 (2013).

29. Bron, P. *et al*. Li$_{10}$SnP$_2$S$_{12}$: An affordable lithium superionic conductor. *J. Am. Chem. Soc.* **135**, 15694–15697 (2013).

30. Whiteley, J. M., Woo, J. H., Hu, E., Nam, K.-W. & Lee, S.-H. Empowering the lithium metal battery through a silicon-based superionic conductor. *J. Electrochem. Soc.* **161**, A1812–A1817 (2014).

31. Seino, Y., Ota, T., Takada, K., Hayashi, A. & Tatsumisago, M. A sulphide lithium super ion conductor is superior to liquid ion conductors for use in rechargeable batteries. *Energy Environ. Sci.* **7**, 627–631 (2014).

32. Lin, Z., Liu, Z., Dudney, N. J. & Liang, C. Lithium superionic sulfide cathode for all-solid lithium–sulfur batteries. *ACS Nano* **7**, 2829–2833 (2013).

33. Murayama, M., Sonoyama, N., Yamada, A. & Kanno, R. Material design of new lithium ionic conductor, thio-LISICON, in the $Li_2S–P_2S_5$ system. *Solid State Ion.* **170**, 173–180 (2004).

34. Li, T. *et al.* A native oxide high-$\kappa$ gate dielectric for two-dimensional electronics. *Nat. Electron.* **3**, 473–478 (2020).

35. Robertson, J. High dielectric constant oxides. *Eur. Phys. J. Appl. Phys.* **28**, 265–291 (2004).

36. Sachid, A. B. *et al.* Monolithic 3D CMOS using layered semiconductors. *Adv. Mater.* **28**, 2547–2554 (2016).

37. Tong, L., Wan, J., Xiao, K. *et al.* Heterogeneous complementary field-effect transistors based on silicon and molybdenum disulfide. *Nat. Electron.* **6**, 37–44 (2023).

38. Kang, W.-M., Cho, I.-T., Roh, J., Lee, C. & Lee, J.-H. High-gain complementary metal-oxide-semiconductor inverter based on multi-layer $WSe_2$ field effect transistors without doping. *Semicond. Sci. Technol.* **31**, 105001 (2016).

39. Koenig, S. P. *et al.* Electron doping of ultrathin black phosphorus with Cu adatoms. *Nano Lett.* **16**, 2145–2151 (2016)

40. Liu, T. *et al.* Nonvolatile and programmable photodoping in $MoTe_2$ for photoresist-free complementary electronic devices. *Adv. Mater.* **30**, 1804470 (2018).

41. Yu, L. *et al*. Design, modeling, and fabrication of chemical vapor deposition grown MoS$_2$ circuits with E-mode FETs for large-area electronics. *Nano Lett.* **16**, 6349–6356 (2016).

42. Wachter, S., Polyushkin, D., Bethge, O. *et al*. A microprocessor based on a two-dimensional semiconductor. *Nat. Commun.* **8**, 14948 (2017).

43. Lei, B. *et al*. Manipulating high-temperature superconductivity by oxygen doping in Bi$_2$Sr$_2$CaCu$_2$O$_{8+\delta}$ thin flakes. *Natl. Sci. Rev.* **9**, nwac089 (2022).

44. Leng, X., Garcia-Barriocanal, J., Bose, S., Lee, Y. & Goldman, A. M. Electrostatic control of the evolution from a superconducting phase to an insulating phase in ultrathin YBa$_2$Cu$_3$O$_{7-x}$ films. *Phys. Rev. Lett.* **107**, 027001 (2011).

45. Lee, P. A., Nagaosa, N. & Wen, X.-G. Doping a Mott insulator: Physics of high-temperature superconductivity. *Rev. Mod. Phys.* **78**, 17–85 (2006).

46. Liao, M. *et al*. Superconductor–insulator transitions in exfoliated Bi$_2$Sr$_2$CaCu$_2$O$_{8+\delta}$ flakes. *Nano Lett.* **18**, 5660–5665 (2018).

## Methods

**Film growth and structural characterization of metal fluoride dielectrics**

The RE-F$_3$ fluoride compound powders (purity: 99.99%, specification: 1–3 mm) were used to evaporate the RE-F$_3$ film onto Si/SiO$_2$ wafers in a thermal evaporation system with a background vacuum better than $10^{-7}$ Torr. The RE-F$_3$ powders purchased from different companies can produce similar properties of fluoride dielectric films (the quality of the fluoride dielectric films does not depend on which company they are originally from). We performed room-temperature deposition of fluoride films to intentionally generate more F$^-$ ion vacancies in fluoride films. The deposition rate was kept at 0.4 Å/s. There was no post-annealing process after film deposition. The thickness and surface morphology of the evaporated fluoride films were confirmed using atomic force microscopy (AFM, integrated with WITec Alpha 300). The powder X-ray diffraction (XRD, Bruker D8 Discover diffractometer $\lambda = 1.5418$ Å) was performed to determine the crystal structure of fluoride films. High-angle annular dark-field scanning transmission electron microscopy (HAADF-STEM) and energy dispersive spectroscopy (EDS) mappings of fluoride films were characterized on an aberration-corrected FEI Titan G2 60-300 STEM at 300 kV with a SuperX EDS system. The convergence semi-angle of the probe was 25 mrad.

**Electric double-layer charging dynamics with electrochemical impedance spectroscopy**

The EDL charging process in fluoride films was investigated using EIS with an electrochemical workstation (Zahner, Zennium Pro). Based on the Au/fluoride/Au parallel plate capacitor geometry (with 200 nm fluoride films), EIS measurements were performed with an applied AC voltage of 20 mV and a frequency $f$ ranging from 10 mHz to 1 MHz to obtain the frequency-dependent impedance $Z$ and phase angle $\theta$. The capacitance ($C$) per unit area of fluoride films was derived from the equation $C = 1/2\pi f Z'' S$, where $f$ is the frequency, $Z''$ is the imaginary part of the impedance, and $S$ is the area of the parallel plate capacitors. The capacitance value at 10 mHz is used to represent the EDL capacitance $C_{EDL}$ of all the Au/fluoride/Au parallel plate capacitors. Temperature-dependent EIS measurements were performed under vacuum conditions in a cryo-free low-temperature system.

**Fabrication of fluoride-gated MoS$_2$ (or WSe$_2$) transistors and Bi-2212 devices**

The $n$-type ($p$-type) MoS$_2$ (WSe$_2$) transistors were fabricated as the prototype transistors device to demonstrate the advance of fluoride dielectrics, as shown in Supplementary Fig. 19. The back-gate electrodes (Ti/Au: 5/20 nm) were deposited onto a Si/SiO$_2$ substrate. The fluoride films were deposited on the back-gate electrodes as the dielectric layer. Few-layer MoS$_2$ (or WSe$_2$) samples were cleaved from bulk crystals onto polydimethylsiloxane by mechanical exfoliation, and the flakes with the desired thickness were then transferred onto the substrate with back-gate electrodes. The final device was fabricated by using an electron-beam lithography process, and source-drain electrodes (Ti/Au: 5/40 nm) were deposited by a metal evaporation process. The logic circuits were constructed by electrically connecting $n$-type MoS$_2$ transistors and $p$-type WSe$_2$ transistors with patterned electrodes for a certain geometry. Bi-2212 flakes with the desired thickness were cleaved from bulk crystals onto PDMS (polydimethylsiloxane) by mechanical exfoliation and transferred to a Si/SiO$_2$ substrate with pre-patterned electrodes. Such pre-patterned (Ti/Au: 3/12 nm) electrodes and a side gate electrode were fabricated using the electron beam evaporation method. To avoid sample degradation, the whole sample preparation was processed in a glove box with nitrogen gas protection.

**Electronic transport measurements**

Electronic transport measurements were performed in a cryo-free superconducting magnet system (Oxford Instruments TeslatronPT). Electrical characterization was carried out under vacuum conditions at room temperature. The DC source-drain voltage and gate voltage were independently applied by two Keithley 2400 sourcemeters. In the $MoS_2$-$WSe_2$ inverter, a Keithley 2182 nanovoltmeter was used to probe the DC output voltage (voltage drop on *p*-type $WSe_2$ transistors). Within the inverter constructed by electrically connecting $MoS_2$ and $WSe_2$ transistors, the gate voltage of the *n*-type $MoS_2$ transistors was used as the input signal ($V_{IN}$), and the voltage drop of the *p*-type $WSe_2$ transistors was used as the output signal ($V_{OUT}$) in the inverter measurements. For high-frequency measurements, a Keysight 33600A waveform generator was used to generate the input signal for fluoride-gated transistors and inverters, and a Tektronix TBS 1202B digital oscilloscope was used to measure the dynamic output voltage.

**First-principles calculations and *ab initio* molecular dynamics simulations**

First-principles calculations are performed using density functional theory (DFT)[47] as implemented in the Vienna *Ab initio* Simulation Package (VASP)[48] within the projector augment wave (PAW) method[49], in which the Perdew-Burke-Ernzerhof (PBE) generalized-gradient approximation (GGA) functionals are used. We used $\sqrt{3} \times \sqrt{3} \times 2$ supercells (including 144 atom sites) with $2 \times 2 \times 2$ gamma centered Monkhorst–Pack *k*-point meshes. The structure model in our DFT calculations and *ab initio* molecular dynamics (AIMD) simulations is based on one fluorine vacancy in each $\sqrt{3} \times \sqrt{3} \times 2$ supercell with 108 total fluorine sites. Such supercell structures were relaxed until the Hellmann-Feynman force on each atom was less than 0.001 eV/Å. The climbing image nudged elastic band (CI-NEB) method is used to predict the energy barriers of different diffusion paths for F⁻ ions.

**Data availability**

The data that support the plots within this paper and other finding of this study are available from the corresponding authors upon reasonable request. Source data are provided with this paper.

**Methods-only References**

47. Kohn, W. & Sham, L. J. Self-consistent equations including exchange and correlation effects. *Phys. Rev.* **140**, A1133–A1138 (1965).

48. Kresse, G. & Furthmüller, J. Efficiency of *ab initio* total energy calculations for metals and semiconductors using a plane-wave basis set. *Comput. Mater. Sci.* **6**, 15–50 (1996).

49. Blöchl, P. E. Projector augmented-wave method. *Phys. Rev. B* **50**, 17953–17979 (1994).